\documentclass[multphys,vecphys]{svmult}

\usepackage{makeidx}         
\usepackage{graphicx}        
\usepackage{multicol}        
\usepackage[bottom]{footmisc}

\makeindex             

\begin{document}

\title*{Probing the Environment with Galaxy Dynamics}
\author{Aaron J. Romanowsky
}
\institute{Departamento de F\'{i}sica, Universidad de Concepci\'{o}n,
Casilla 160-C, Concepci\'{o}n, Chile
\texttt{romanow@astro-udec.cl}
}
%
%
\maketitle

\begin{abstract}
I present various projects to study the halo dynamics of elliptical galaxies.
This allows one to study the outer mass and orbital distributions of ellipticals in different 
environments, and the inner distributions of groups and clusters
themselves.
\end{abstract}

\section{Introduction: Halos and the environment}
\label{romanow:sec:1}
Elliptical galaxies are intriguingly homogeneous.
Besides being more prevalent in high density regions, 
low-redshift ellipticals have observable central properties
(e.g. color, velocity dispersion, metallicity, star formation history)
whose environmental dependencies are relatively subtle
(e.g. \cite{romanow:reda05,romanow:clemens}), in contrast to the case of spiral galaxies.
Larger differences might be found in the ellipticals' outer parts,
which should be the most strongly affected by environmental influences.
For example, the tidal fields in higher-density environments
could cause stripping of the galaxies' stellar and dark matter (DM) halos, and of their
globular cluster (GC) systems---also affecting the anisotropy of the remaining
halo stars and GCs.
Indeed, 
a deep Virgo Cluster image seems to show stellar halo stripping in progress, a
process which may be facilitated
by dynamical halo heating in the pre-infall groups \cite{romanow:mihos}.
Gravitational lensing studies also indicate some DM halo stripping
in high-density environments
\cite{romanow:natarajan02,romanow:gavazzi04,romanow:natarajan04,romanow:mandelbaum06}.

The large radial extent of GC systems makes them handy tracers for halo stripping,
and in fact it may be possible to use GCs as a proxies for the DM itself
\cite{romanow:bekki05}.
Very extended GC systems are found in
cluster-dominant ellipticals (such as M87, M49, and NGC~1399)
\cite{romanow:mclaughlin99},
which is consistent with them being agents rather than victims of stripping.
Wide-field studies of more normal ellipticals are now getting underway---and
provocatively, the Virgo galaxy NGC~4636 turns out
to have a GC system with a sharp edge \cite{romanow:dirsch05}.

Groups should dominate the error budget of the Universe, but their mass
distributions are among the most poorly determined.
Observations from weak lensing and internal group dynamics
don't agree on their total masses,
much less the detailed distribution with radius
(e.g. \cite{romanow:carlberg01,romanow:tully05,romanow:kara05,romanow:parker05}).
Additional constraints, especially nearer the group centers, are essential.
Fortunately, studies of elliptical galaxy halos
allow one to probe the
mass distributions in galaxy groups as well as in individual galaxies.
This is because many of the easiest-studied ellipticals 
are central group or cluster galaxies
(e.g. \cite{romanow:fuk06,romanow:kochanek06,romanow:humphrey06})---a frustration
for galaxy research but a boon for group studies.

Whether in galaxies or in groups, there are several useful independent dynamical tracers:
GCs, X-ray gas, and planetary nebulae (PNe), which are
a powerful proxy for faint starlight.
In groups with only a handful of galaxy velocities,
there may be hundreds or thousands of GC and PN velocities attainable
(albeit at relatively small radii).
In M49 and M87, massive DM halo cores are found from
constant-to-rising profiles of gas temperature, and of velocity dispersion
profiles of PNe and GCs \cite{romanow:romanowsky01,romanow:matsushita02,romanow:cote03,romanow:humphrey06,romanow:misc}.
Different orbital properties are implied between the PNe and GCs,
which should constrain central galaxy evolution scenarios.
NGC~4636 has a constant GC dispersion profile, implying a fairly normal
DM halo \cite{romanow:schuberth06}---unsurprisingly different from X-ray results, given
the evident departures from gas equilibrium
\cite{romanow:n4636}.
NGC~1399 has a rising PN and GC dispersion profile, indicating either the
DM core of the Fornax Cluster, or a recent interaction with another galaxy
\cite{romanow:nap02,romanow:n1399}.
Note that for halo tracers to be fruitful, it is imperative to combine them with
constraints on the central galaxy's stellar mass, which is otherwise a major
source of systematic uncertainty \cite{romanow:humphrey06}.

\section{The Eridanus~A Group}
\label{romanow:sec:eri}

\begin{figure}
\centering
\includegraphics[height=8cm]{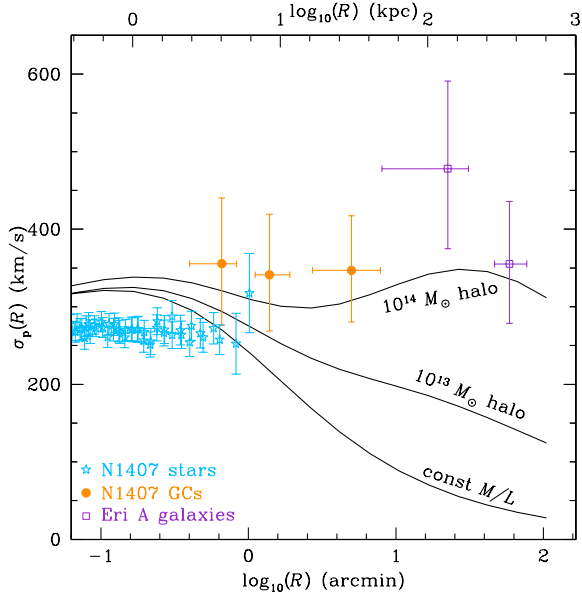}
%
%
\caption{
Projected velocity dispersion radial profiles in the Eridanus~A group.
Points with error bars show
 data for stars and GCs in NGC~1407 \cite{romanow:n1407}, and for group galaxies.
Curves show model predictions for the GCs, for a spherical isotropic assumption,
and either no dark matter, a ``normal'' halo, or a ``super'' halo.
}
\label{romanow:fig:1}       
\end{figure}

Based on its X-ray gas and member galaxy velocities, Eridanus~A appears to be
a ``dark cluster'' with a virial mass $\sim 10^{14} M_{\odot}$ and
mass-to-light ratio  $\Upsilon_B \sim$~1500~$\Upsilon_{B,\odot}$ \cite{romanow:gould93,romanow:quintana94,romanow:humphrey06,romanow:brough}.
To probe this possibility further,
we have acquired velocities of $\sim$~100 GCs around
the central giant elliptical NGC~1407, using
LRIS, FLAMES, and LDSS-3.
Initial results with 36 GCs, comparing the GC dispersion profile to various models,
 do indeed support the presence of a super-halo
(see Fig.~\ref{romanow:fig:1}).
Based on typical empirical and theoretical values for the virial $\Upsilon$ of a group like
Eri~A
\cite{romanow:vdB03,romanow:parker05,romanow:eke06},
the profile should decrease outside 10 kpc---but
a flat dispersion is observed.
Combining more constraints from stellar, GC, X-ray, and group galaxy dynamics will allow us
to trace the mass profile in more detail.

\section{The Leo~I Group}
\label{romanow:sec:leo}

\begin{figure}
\centering
\includegraphics[height=8cm]{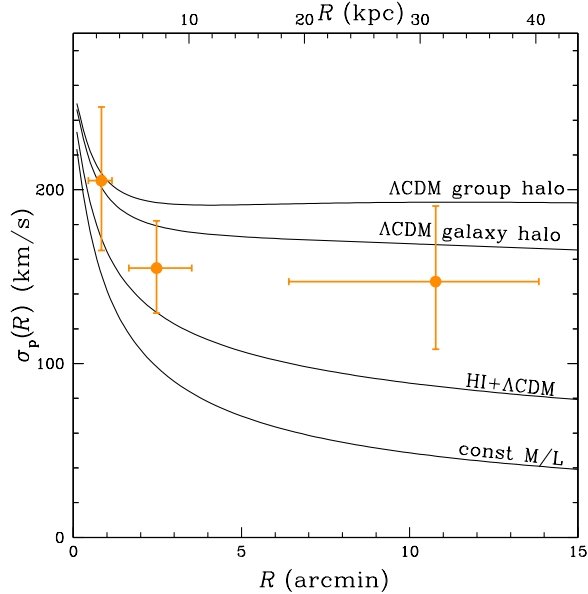}
%
%
\caption{
Dispersion profile of GCs in the Leo~I group.
Error-bars show the data, and curves show spherical isotropic model predictions.
The typical early-type galaxy halo mass comes from weak lensing results \cite{romanow:mandelbaum06},
and group mass from luminosity-function studies \cite{romanow:vdB03,romanow:eke06}.
A halo consistent with the HI ring constraint is also shown \cite{romanow:schneider85}.
}
\label{romanow:fig:2}       
\end{figure}

Leo~I is the nearest (10 Mpc) example of a group containing
multiple giant early-type galaxies, including the ``archetypal''
$L^*$ elliptical at its center, NGC~3379.
It is unclear if the group is reasonably relaxed, with a group halo centered on NGC~3379.
This galaxy has been the focus of numerous dynamical studies, most
recently employing PNe \cite{romanow:romanowsky03,romanow:sluis06} and GCs \cite{romanow:puzia04,romanow:pierce06,romanow:bergond06}.
The PNe imply surprisingly little DM inside 15~kpc, but leave the possibility
that there are large amounts of DM spread further out in a massive, diffuse halo.
The GC kinematics reach to 40 kpc, and do suggest a massive, even group-sized, halo 
(see Fig.~\ref{romanow:fig:2}).
Another remarkable constraint comes from the HI gas ring which appears to
orbit the core of the group, and implies $\Upsilon_B \sim$~30 inside 100 kpc \cite{romanow:schneider85}.
This suggests a lot of DM, but much less than one would expect for a $\Lambda$CDM
halo (whether galaxy- or group-sized), and appears to be at odds with the GC results
(the PNe are technically compatible with either the GCs or the HI).
However, the GC constraints are hampered by
small-number statistics (49 velocities), and newly-acquired
FLAMES data should clarify the situation with a doubled or tripled data set.
As an interesting note of comparison, Leo~I and Eri~A are groups with nearly the same optical
luminosity, but apparently differ in mass by at least a factor of 10.

\section{Mass profiles of ordinary ellipticals}
\label{romanow:sec:prof}

The halo mass profiles of ``ordinary'' ($\sim L^*$) ellipticals have
long been elusive.
The first inroads have come from new data on halo PN kinematics
\cite{romanow:mendez01,romanow:romanowsky03,romanow:teodorescu05}.
The first 5 ellipticals studied all have a projected velocity dispersion profile
which declines markedly with radius.
The obvious implication is that the measurable DM content of these galaxies is
remarkably low, as discussed above for NGC~3379.
One might expect this effect to be strongest for the highest-density environments,
where DM stripping could have occurred, but this does not appear to be the case.
An alternative possibility is that the galaxy halo concentrations are lower
than expected with $\Lambda$CDM, a possibility strengthened by analysis of
literature data on early-type galaxies \cite{romanow:napolitano05}.
This is supported by some independent studies of ellipticals
\cite{romanow:borriello03,romanow:rusin03,romanow:fuk06}, and is paralleled by many studies of late-type galaxies.

Theoretical studies \cite{romanow:mamon05,romanow:dekel05} have pointed out various
effects which could contribute to the declining dispersions,
including oversimplified DM profiles, radial anisotropy variations,
galaxy flattening, and biased PN-stellar correspondence.
There are reasons to doubt that these effects could entirely explain
the observations---for example, the modeling of NGC~3379 already incorporated a direct
derivation of the (radial) anisotropy profile from the data.
But certainly much more clarification is needed.
Current work focuses on obtaining a large, systematic sample of independent
mass tracers around ordinary ellipticals in different environments, and 
on refining the modeling---including direct
calibratory comparisons with simulations.

%
%
%

%
%



\printindex
\end{document}